\documentclass[a4paper]{article}
\usepackage[top=50pt,bottom=60pt,left=70pt,right=60pt]{geometry}
\usepackage[affil-it]{authblk}
\usepackage[sectionbib,square]{natbib}
\usepackage{graphicx}
\usepackage{eucal}
\usepackage{amsxtra}
\usepackage{comment}
\usepackage{color}
\usepackage{abstract}

\usepackage{hyperref}
\hypersetup{
  pdfauthor={V.Kulikovskiy},
  pdftitle={Adding a systematic uncertainty to the signal estimation in the on/off-zone measurements.},
  pdfsubject={Statistics},
  urlcolor=blue,
}

\title{Adding a systematic uncertainty to the signal estimation in the on/off-zone measurements.}
\author{Vladimir Kulikovskiy
\thanks{vladimir.kulikovskiy@ge.infn.it}}
\affil		{
\footnotesize APC, Universit\'e Paris Diderot, CNRS/IN2P3, CEA/IRFU, Observatoire de Paris, \\Sorbonne Paris Cit\'e, 75205 Paris, France\\
		INFN Sezione di Genova, via Dodecaneso 33, Genoa, Italy \\
                 Universit\`a Degli Studi di Genova, via Dodecaneso 33, Genoa, Italy \\
                 SINP MSU, Vorobyevi gory, 1, Moscow, Russia
                 }    
 
 \date{Dated: \today}

\begin{document}
\maketitle
\begin{abstract}
{\it
The measurements with the background estimation from an off-zone are widely used in astrophysics and accelerator physics. 
In this note an overview of the statistical methods which estimate the range and the significance of the measured signal is done. The method which includes a systematic uncertainty is developed for the on/off-zone measurements and compared with other existing methods.}
\end{abstract}

\section{Introduction}
The main goal of any signal measurement is to estimate the signal value and the precision of the estimation. For this purpose the most probable value of the signal and also the the signal range are provided as a results of the measurement. The probability of the signal to be between the range limits is fixed to some value (90\% for example or $1-\alpha$). In terms of statistics this mean that measurement $X$ was done and one wants to provide the value of the signal $\hat s$ which maximises the probability $P(s \mid X)$ and also the limits $s_{min}$, $s_{max}$ for which $P(s_{min}<s<s_{max})=1-\alpha$. The values $\hat s$, $s_{min}$, $s_{max}$ can be calculated if $P(s \mid X)$ is known for every $s$. 

Actually in the most cases it is straightforward to know the probability $P(X \mid s)$ of observing $X$ if the signal $s$ is known. This can be done assuming Poisson or another model involved in the measurements. To obtain $P(s \mid X)$ one can use the Bayes' theorem:
\begin{equation}
P(s \mid X)=\frac{P(X \mid s)P(s)}{P(X)}
\label{formula:1}
\end{equation}

For $P(s \mid X)$ the denominator is simply a constant as it depends only on the measurement $X$, so it can be calculated using the probability property $\int_{-\infty}^{+\infty} P(s \mid X)\,\mathrm{d}s =1$. The remaining multiplier $P(s)$ is the unknown probability distribution of the signal. This formula shows that strictly saying the measurement $X$ does not give the answer on how much could be the signal. To go beyond this paradox two approaches were developed by the statisticians. As the "true limits" cannot be obtained it is expected at least to have some reasonable estimations which should follow na\"ive expectations such as:
\begin{itemize}
\item The upper limit is smaller if the measured value is smaller in the measurements which are done with the same background.
\item The limits are broader if the background is higher in the measurements with the equally estimated value of the signal $\hat s$.
\item The upper limit is lower if the systematic errors are smaller in the measurements with the same background and the same measured $X$.
\end{itemize}

To find $P(s \mid X)$ from the formula (\ref{formula:1}) one can use a $P_{theo}(s)$ which is based on some belief, physical limits or previous measurements. This approach is called Bayesian. The simplest estimation of $P_{theo}(s)$ may be realised by setting a uniform distribution between the physical limits. However it is known that the limits are different for the different $P_{theo}(s)$. $P(s\mid X)$ obtained from the formula~(\ref{formula:1}) using $P_{theo}(s)$ is named a post probability. 

An another approach is used by the frequentists. In this approach one sets limits in a way to have a true value of the signal inside the limits in $1-\alpha$ part of the all measurements. This is also called a coverage of the method. The advantage on this approach is that it is based only on $P(X \mid s)$ or, equivalently, on the likelihood $\mathcal{L}(s \mid X)\equiv P(X \mid s)$\footnote{Please, note that in the following, the order of the observables is changed with the variables in the likelihood description, so $\mathcal{L}(X \mid s)\equiv\mathcal{L}(s\mid X)$. This is convenient to describe the likelihood with classical processes/distributions like $\text{Gaussian}(x \mid \tilde{x},\sigma)$ and $\text{Poisson}(x \mid \lambda)$ where the variables (which are normally the observables and the nuisance parameters of an experiment) are separated from the parameters (which are normally the known or searched variables of an experiment). Integration of the likelihood/probability over all sets in the left part from the "$|$" should give unity.}.

 A frequentist method developed by Neyman consists of the two steps~\citep{Neyman}:
\begin{itemize}
\item To find some "confidence belt" for every signal $s$. The confidence belt is a set of different measurements $\{X_{n}^s\}$ which give $\sum_{i=1}^{n}\mathcal{L}(s \mid X_{i}^s)\ge1-\alpha$. The way to choose the confidence belt $\{X_{n}^s\}$ is arbitrary. 
\item To use a real measurement $X$. If $X$ is present in some set of $\{X_{n}^s\}$ for the signal $s$ then this signal $s$ is included to the confidence area. This operation is done for every set $\{X_{n}^s\}$ to build the area. The upper limit with $1-\alpha$ confidence level is the maximum signal in the confidence area.
\end{itemize}

Limits obtained by this method may vary depending on the choice of the confidence belts. In~\citep{Neyman} two options were suggested: central and one-sided limits (upper/lower). Both types have some unwanted features. Central limits may be too strict for the case when the number of the observed events is lower than the expected number from the background. One-sided limits do not allow to claim a discovery or setting the upper limit depending on the measured $X$. The mentioned problems are essential as one should define which type of the limits one wants to set before the measurement. If this decision is done after the measurement this would lead to the under-coverage of the limits known as a "flip-flopping" problem. 

\section{Case of the signal with the known background}
The methods described here are applied to the processes with a Poisson likelihood:
\begin{equation}
\label{poisson}
     \mathcal{L}(n_\text{obs}\mid s;b)=\frac{s^{n_\text{obs}}e^{-s}} {n_\text{obs}!}\equiv \text{Poisson}(n_\text{obs}\mid s+b) 
\end{equation}
where $X=n_\text{obs}$ and ";" symbol is used to distinguish the unknown variables from the known ones which are for this case the signal and the background correspondingly.

In order to avoid the "flip-flopping" problem the Neyman's method was improved by the special ordering principle~\citep{FC}. The advantage is that the procedure to define an upper limit in case of a low number of observed events or a two-sided limit in the opposite case is defined {\it a priori}. Confidence belt selection is based on sorting the $\{X_{n}\}$ by the decreasing rank:
\begin{equation}
\label{ratio}
     R = \frac{\mathcal{L}(n_{obs} \mid s;b)}
                   {\mathcal{L}(n_{obs} \mid \hat{s};b)}
\end{equation}
where $\mathcal{L}(n_\text{obs}\mid \hat{s};b)$ is a maximum possible likelihood for the fixed number of observed events $n_\text{obs}$ which is realised by the signal $\hat{s}$.

For some cases of the low number of the observed evens the measurements with a higher background may provide better limit with this method. Actually, this seeming discomfort is connected with the difference between Bayesian and frequentist approaches. In the latter, if a signal plus background is measured, then if the expected mean background is bigger the signal should be lower (in average over a lot of trials). But, on the another hand, it is expected that the experiment with bigger expected background can not provide better limits. 
 
 Bayesian method consists of the following steps:
 \begin{itemize}
 \item To choose $P_{theo}(s)$. In a simple case a uniform prior distribution may be chosen. In this case the post probability $P(s)\propto\mathcal{L}(n_\text{obs}\mid s;b)$. Some sophisticated approaches may have a more complicated prior, but the obtained limits are similar in the most cases.
 \item To find $s_\text{lower}$ and $s_\text{upper}$ using $n_\text{obs}$ from the measurement. The integration of the likelihood till $1-\alpha$ confidence level is used:
\begin{equation}
\frac{\int_{s_\text{lower}}^{s_\text{upper}}\mathcal{L}(n_\text{obs}\mid s;b)\,\mathrm{d}s}
{\int_{-\infty}^{\infty}\mathcal{L}(n_\text{obs}\mid s;b)\,\mathrm{d}s} \ge 1-\alpha
\end{equation}
The integration can be done numerically using steps for $s$ and ordering the likelihood corresponding to each step by the decreasing order. Once the ratio becomes equal or more than $1-\alpha$ the integration stops and the minimum and maximum values of $s$ become the limits.
 \end{itemize}
 
\begin{figure}
\begin{center}
\includegraphics[width=0.8\textwidth]{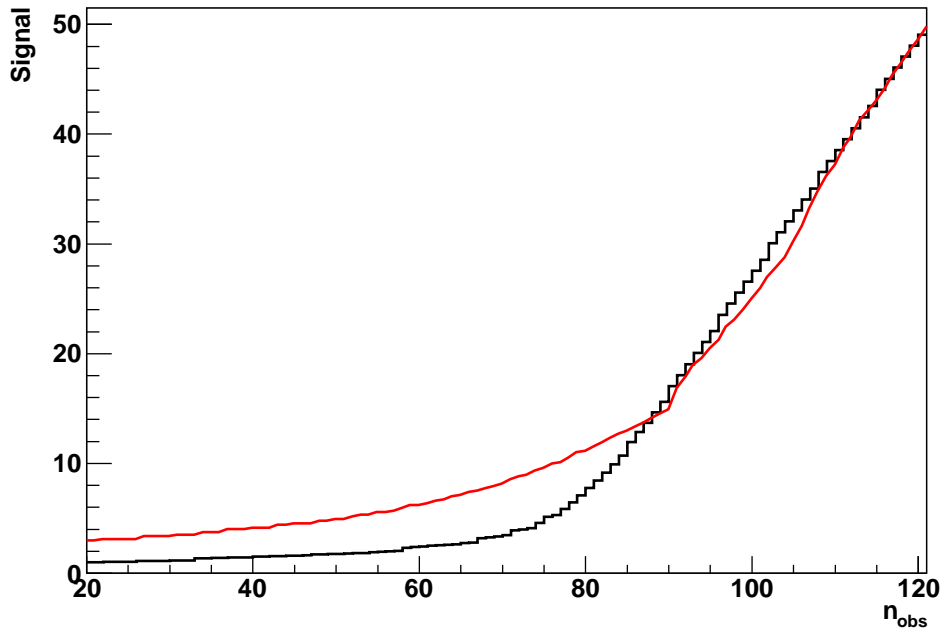}
\caption{Calculation of the upper limits with the Feldman \& Cousins universal approach (solid line) and the Bayesian approach (dotted line) for the case of $90$ expected background.
}
\label{fig:bayesfc}
\end{center}
\end{figure}

 Comparison between the limits obtained by the Feldman \& Cousins universal approach and the Bayesian approach is presented in fig.~\ref{fig:bayesfc}. The limits are very similar for the case when $n_{obs}>b$ and the frequentist upper limit is tighter when $n_{obs}<b$. The latter is a known problem which was discussed a lot during last two decades. As a solution some mixed approaches were developed.
 
One of the proposed solutions was in a usage of a so-called conditional PDFs~\citep{Roe,Conrad}. This PDF uses the current measurement $n_{obs}$ and the fact that the background can not be bigger than the $n_{obs}$. The advantage is that for zero observed events upper limit is about 2.42 independently from the background mean $b$. Instead, the universal approach provides upper limit 2.44 for zero mean background, 1.08 for 3 mean background and so on decreasing, in contrast to the Neyman's upper limit which is 2.3 for zero background. Lately it was found that this method with a conditional PDFs suggests a presence of the signal in all cases if the observed value is bigger than the mean background for the continuous observable~\citep{Cousins}. Also the method does not have a coverage in the frequentist sense. One of the proposed solutions was to change the lower limit to have $1-\alpha$ coverage for each signal~\citep{Mandelkern}. Another solution, proposed by the authors of the conditional PDF themselves was to use a Bayesian limits with the uniform distribution of the signal in the physically allowed range ($>=0$)~\citep{Roe2}.

Actually, even acknowledging this discomfort with an upper limit decrease while increase of the background in case of zero events Feldman \& Cousins method is still a default method for the upper limits calculation among physicists. However in the form it is described in the former paper it does not allow to include systematic errors in the calculation. 

\section{Including uncertainties to the known background}
For a measurement with a background with a known uncertainty $b\pm\sigma_{b}$ (statistic plus systematic error) the likelihood can be modified as:
\begin{equation}
\label{lihelihoodsysstd}
     \mathcal{L}(n_\text{obs}, b' \mid s;b,\sigma_b) = \text{Poisson}(n_{obs}\mid s+b')\text{Gaussian}(b'\mid b, \sigma_{b})
\end{equation}

where $b'$ is the mean background in the measurement $n_{obs}$. It is assumed that $b'$ has a gaussian distribution with a mean $b$ and a sigma $\sigma_{b}$ (however in some literature systematic errors are approximated with a uniform distribution). Actually the presence of the distribution of $b'$ in this likelihood referred to a Bayesian approach. So, the limits obtained by using this likelihood are Bayesian or semi-Bayesian depending on the further calculation procedure. 

Investigating the tools available in the ROOT framework the method described by Rolke {\it et al }~\citep{Rolke} was found. This method is a modified "$\mathcal{L}+\frac{1}{2}$" method which means that it is a maximum likelihood estimator (MLE) where the likelihood ratio is used to find the most probable signal and the background. Practically, MLE method is identical to the Bayesian approach with the uniform prior distribution of the signal and the profile likelihood $\mathcal{L}(n_\text{obs}, \hat{b'}\mid s,b,\sigma_b)$. Here $\hat{b'}$ means that for every $n_{obs}$ the best value of $b'$ which maximises the likelihood is found. The obtained likelihood does not depend on $b'$ anymore.

\begin{figure}
\begin{center}
\includegraphics[width=0.8\textwidth]{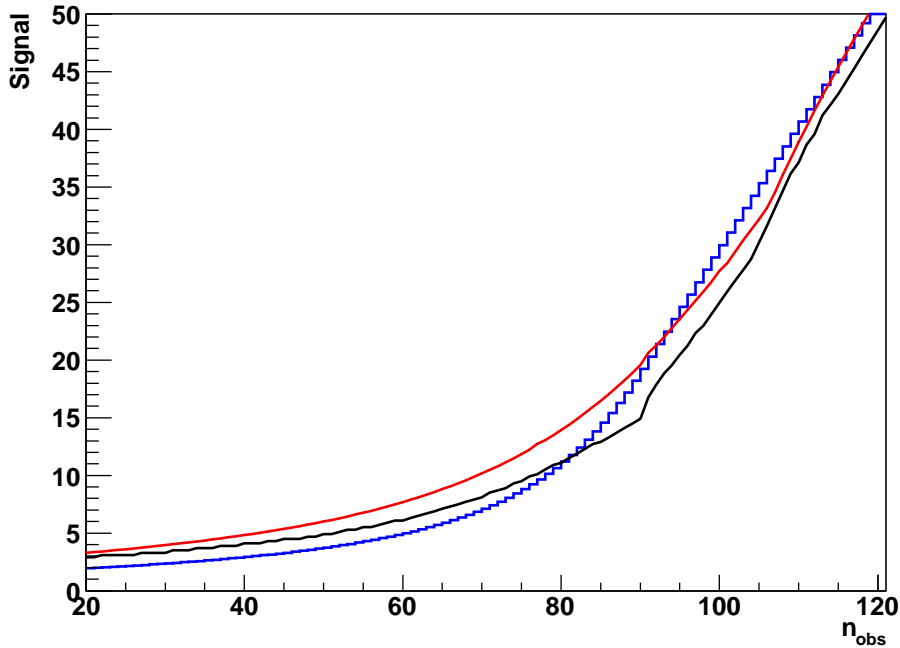}
\caption{Upper limit calculation with the Bayesian method using the profile likelihood~(\ref{lihelihoodsysstd}) and $\chi^2$ approximation~\citep{Rolke} (blue) compared to the calculation with the exact Bayesian method with the same profile likelihood (red) for the case of $90\pm6$ expected background. The upper limit with the Bayesian method for the case of the $90$ expected background~(\ref{poisson}) is also shown (black).
}
\label{fig:rolke_bayes_fc}
\end{center}
\end{figure}

As in the Bayesian approach limits $s_{lower}$, $s_{uppper}$ according to~(\ref{formula:1}) should satisfy the equation:
\begin{equation}
\frac{\int_{s_\text{lower}}^{s_\text{upper}}\mathcal{L}(n_\text{obs}, \hat{b}'\mid s;b,\sigma_b)P_\text{theo}(s)\,\mathrm{d}s}
{\int_{-\infty}^{\infty}\mathcal{L}(n_\text{obs}, \hat{b}'\mid s;b,\sigma_b)P_\text{theo}(s)\,\mathrm{d}s} \ge 1-\alpha
\end{equation}
To make calculations faster the theorem that $-2\log R$ has approximately $\chi^2$  was used~\citep{Rolke}. Exact Bayesian profile likelihood method and the $\chi^2$ assumption is compared in fig.~\ref{fig:rolke_bayes_fc} for the case of $90\pm6$ expected background. It was found that approximation with $\chi^2$ gives lower limits than the exact calculation. Bayesian limit for the background without uncertainty is also presented for the comparison. This limit is always lower than the limit with an uncertainty as it is expected.

Another approach is a frequentist construction of the Neyman's confidence belts based on the so-called marginal likelihood which is a likelihood integrated over $b'$~\citep{Conrad}:
\begin{equation}
\label{f:likstdmarg}
 \mathcal{L}(n_\text{obs} \mid s;b,\sigma_b) =  \int_{0}^{\infty} \mathcal{L}(n_\text{obs},b' \mid s;b,\sigma_b)\,\mathrm{d} {b'}
\end{equation}

\begin{figure}
\begin{center}
\includegraphics[width=0.8\textwidth]{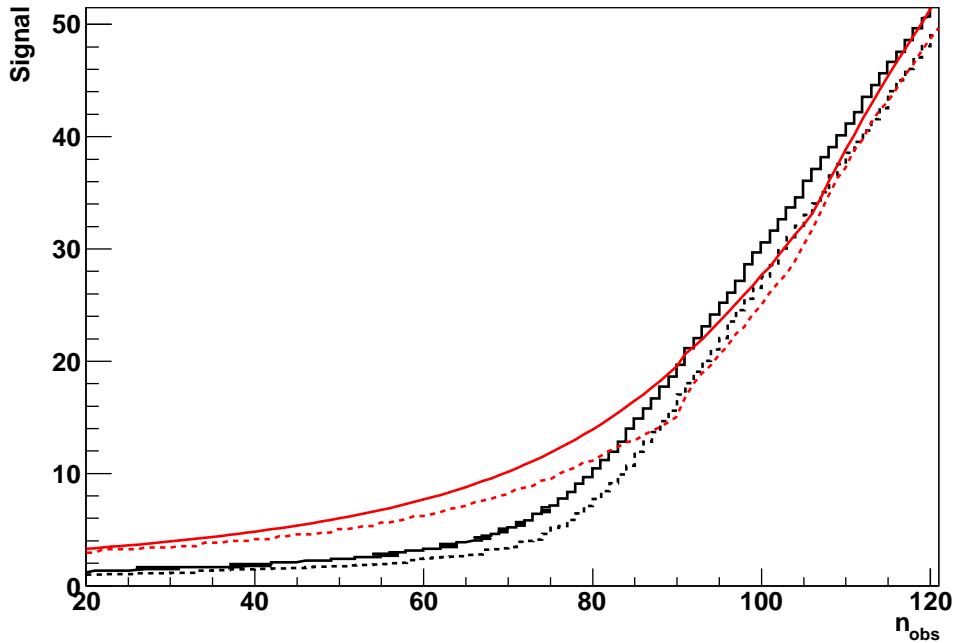}
\caption{Calculation of the upper limit with the Bayesian method using the profile likelihood from~(\ref{lihelihoodsysstd}) (red solid), compared with the upper limit calculated with the frequentist approach using the marginal likelihood~\citep{Conrad} (black solid) for the case of the $90\pm6$ expected background together with the Bayesian upper limit (red dotted) and the limit obtained with the Feldman \& Cousins approach (black dotted) for the case of the background 90 without a systematic error.}
\label{fig:limits_std}
\end{center}
\end{figure}

The obtained likelihood is a function of the signal and the observed events only so the upper limit calculation may follow the Feldman \& Cousins method. The obtained limit for $90\pm6$ expected background in comparison with the Bayesian profile likelihood limit calculation is present in fig.~\ref{fig:limits_std}. Again one can see that the Bayesian limits are higher than the frequentist ones when $n_\text{obs}<b$.

It is worth to note that this approach~\citep{Conrad} has some unwanted behaviour for the measurement with the systematic uncertainty of the signal efficiency. The limits are becoming better while the uncertainty on the efficiency is increasing. The suggested correction was to change the likelihood in the denominator of the rank on the best likelihood given no uncertainty on the efficiency~\citep{Hill}.  For the problem with the uncertainty on the background the original method~\citep{Conrad} behaves correctly so no such correction is mandatory.

\section{On/off-zones measurements}
\label{sec:off}
In many experiments it is possible to estimate the background from an additional measurement in the zone where no signal is expected. This so called off-zones approach is widely used in astrophysics and in accelerator physics. The selected off-zone can be an another part of the sky in the case of the astrophysic measurements, an another part of the energy spectrum or a measurement with the switched off accelerator in the case of the accelerator experiments. It is important to mention that the off-zone can have a bigger statistic, for example the area can be geometrically bigger or the exposure time can be longer. Ratio of the statistic in the off-zone to the on-zone is called $\tau$ in following. So, if the number of the background events in the on-zone is expected to be $b$ then in the off-zone the number of the background events is expected to be $\tau b$. 

During such measurements $n_\text{obs}$ and $n_\text{bg}$ in the on-zone and in the off-zone correspondingly are obtained. The number of the events $n_\text{obs}$ is distributed around $s+b$ and $n_\text{bg}$ is distributed around $\tau b$. The choice of the distribution depends on the experiment. In most cases, a Poisson distribution may be used. So, $n_\text{obs}$ and $n_\text{bg}$ are distributed as:
\begin{equation}
\label{poissnobs}
P(n_\text{obs}\mid s,b) = Poisson(n_\text{obs} \mid b+s)
\end{equation}
\begin{equation}
\label{poissbg}
P(n_\text{bg}\mid b;\tau) = Poisson(n_\text{bg} \mid \tau b)
\end{equation}
And if the two measurements are independent then:
\begin{equation}
\label{poissnobsnbg}
P(n_\text{obs},n_\text{bg} \mid s, b; \tau) = \text{Poisson}(n_\text{obs} \mid b+s)\text{Poisson}(n_\text{bg} \mid \tau b)
\end{equation}

This probability was used for the significance estimation in the gamma-ray astronomy~\citep{LiMa}. Also it might be used for the limits estimation. It is important to note that $n_\text{obs}$ and $n_\text{bg}$ are treated in the formula in a similar way which corresponds to the same measurement procedure for these variables. This likelihood has two parameters $s$ and $b$.

Sometimes, during the on/off-zone measurements size or efficiency of each zone can be estimated with a some limited precision resulting as a systematic error. This fact should be included in the probability distribution. It can be done with a two gaussian distributions. The likelihood distribution in this case becomes as follows:
\begin{multline}
\label{poissnobsnbgeps}
\mathcal{L}(n_\text{obs},n_ \text{bg},\alpha_\text{on},\alpha_\text{off} \mid s,b;\tau,\sigma) = \\
=\text{Poisson}(n_\text{obs} \mid \alpha_\text{on}b+s)\text{Gaussian}(\alpha_\text{on}\mid1,\sigma)\times\\
\times \text{Poisson}(n_\text{bg} \mid \alpha_\text{off}\tau b)\text{Gaussian}(\alpha_\text{off}\mid 1, \sigma)
\end{multline}
where $\sigma$ is a systematic uncertainty.

Analytical/numerical construction of the Neyman intervals with the likelihood ratio ranking is possible if the likelihood is first marginalised over $\alpha_\text{on}$ and $\alpha_\text{off}$. The marginal likelihood becomes a function of $s$ and $b$ variables. The obtained confidence area will depend on the $s$ and $b$ and so the limit on $s$ from this area will be quite conservative. Additionally, the integration over $\alpha_\text{on}$ and $\alpha_\text{off}$ can be done only numerically which makes the computation very heavy.

The Bayesian approach is, instead, easily applicable to this likelihood. The fastest way is to use the profile likelihood which depends only from $s$. The post probability area of $1-\alpha$ confidence level size can be calculated as:
\begin{equation}
\frac{\int_{s_\text{lower}}^{s_\text{upper}}\mathcal{L}(n_ \text{obs}, n_\text{bg},\hat\alpha_\text{on}\hat\alpha_\text{off} \mid s,\hat b;\tau,\sigma)\,\mathrm{d}s}
{\int_{-\infty}^{\infty}\mathcal{L}(n_\text{obs}, n_\text{bg},\hat\alpha_\text{on}\hat\alpha_\text{off} \mid s,\hat b;\tau,\sigma)\,\mathrm{d}s} \ge 1-\alpha
\end{equation}

It was not succeeded to calculate the profile likelihood from~(\ref{poissnobsnbgeps}) analytically. However, from the three equations based on the fact that partial derivatives by $b$, $\alpha_\text{on}$, $\alpha_\text{off}$ are equal to zero at the maximum of the likelihood one can obtain:
\begin{equation}
\alpha_\text{on}(1-\alpha_\text{on})+\alpha_\text{off}(1-\alpha_\text{off})=0
\label{formula:alpha}
\end{equation}
\begin{equation}
(\alpha_\text{on}+\alpha_\text{off}\tau)b^2-(n_\text{on}+n_\text{off}-\frac{s}{\alpha_\text{on}}(\alpha_\text{on}+\alpha_\text{off}\tau))b - \frac{s}{\alpha_\text{on}}n_\text{off}=0
\label{formula:b}
\end{equation}
For a one chosen parameter $\alpha_\text{on}$, for example, other parameters, $b$ and $\alpha_\text{off}$ can be calculated by solving these quadratic equations. If one of the solutions is negative, then zero value should be used instead it. The maximum likelihood can be found by scanning $\alpha_\text{on}$ in a range $1\pm3\sigma$ numerically. The performed calculations with a 1\% precision takes about $\sim1$ sec to find a limit numerically for a single measurement using a 2GHz CPU.

The profile likelihood of~(\ref{poissnobsnbg}) has $b$ value which can be calculated similarly to~(\ref{formula:b}):
\begin{equation}
(1+\tau)b^2-(n_\text{on}+n_\text{off}-s(1+\tau))b-n_\text{off}s=0
\end{equation}

\section{Comparison of the methods}
All these methods were applied to a particular astrophysical measurement for a test. In this measurement three similar off-zone regions with a size of the on-zone were selected on the sky. All four zones (one on-zone and three off-zones) have the same visibility.  The number of the background events expected in the on-zone and the off-zones was estimated to be the same inside a systematic uncertainty of 3\%. After the optimisation of the cuts $n_\text{bg}=270$ events in the three off-zones were seen.

\begin{figure}
\begin{center}
\includegraphics[width=0.8\textwidth]{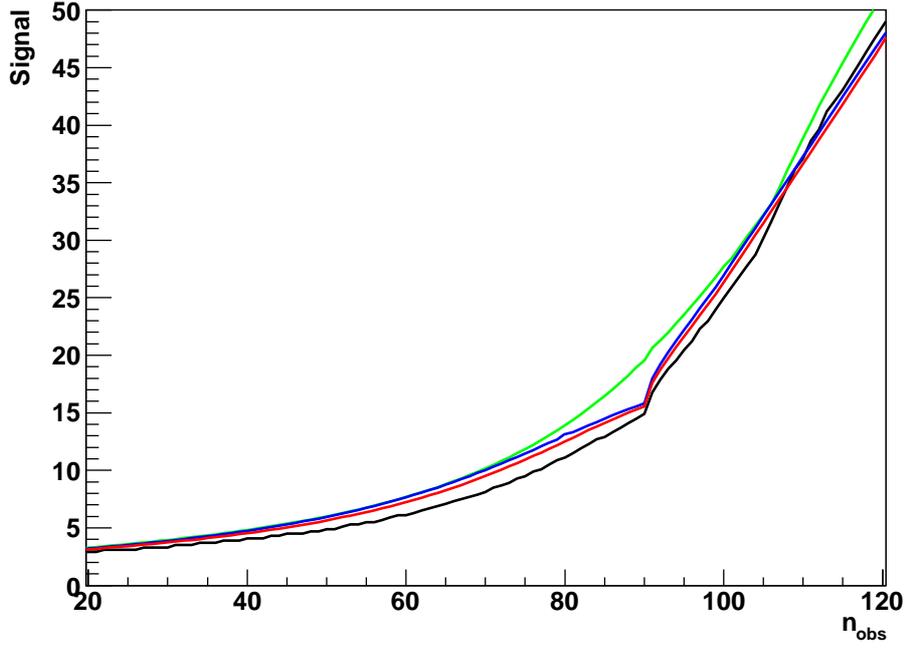}
\caption{Calculation of the upper limits with the Bayesian methods. A process with a known background~(\ref{poisson}) for $b=90$ (black), using a profile likelihood for the process with a known background and a gaussian uncertainty~(\ref{lihelihoodsysstd}) for $b=90$, $\sigma_b=6$ (green), using a profile likelihood for a process with a two Poisson distributions~(\ref{poissnobsnbg}) for $n_{bg}=270$, $\tau=3$ (red) and using a profile likelihood  for a process with a two Poisson distributions with Gaussian uncertainties~(\ref{poissnobsnbgeps}) for $n_\text{obs}=270$, $\tau=3$ and $\sigma=3\%$ (blue).}
\label{fig:all_bayes}
\end{center}
\end{figure}

All Bayesian methods with a different likelihoods are compared in fig.~\ref{fig:all_bayes} for this measurement. Both likelihoods with an introduced background uncertainty provide higher upper limits comparing with a Poisson processes without uncertainty. Likelihood with two Poissons for $n_\text{obs}$ and $n_\text{bg}$~(\ref{poissnobsnbgeps}) provides better limits and follows the shape of the upper limits in the case of the no uncertainty. 

\begin{figure}
\begin{center}
\includegraphics[width=0.8\textwidth]{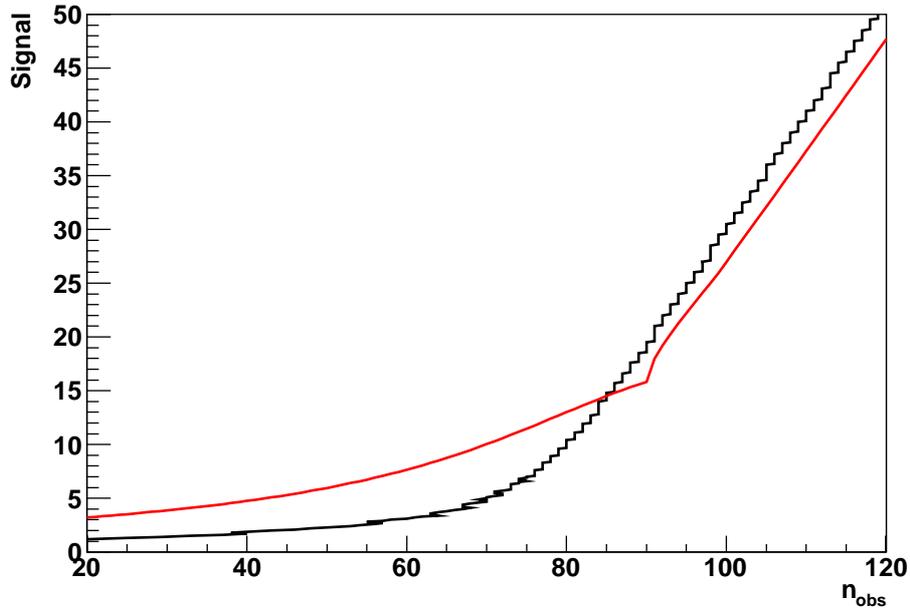}
\caption{Calculation of the upper limits with the likelihood~(\ref{lihelihoodsysstd}) for $b=90$, $\sigma_b=6$ with the Conrad's method (black) and with the likelihood~(\ref{poissnobsnbgeps}) for $n_\text{bg}=270$, $\tau=3$ and $\sigma=3\%$ with the Bayesian method using a profile likelihood (red).}
\label{fig:all_limits}
\end{center}
\end{figure}

The Bayesian approach using a profile likelihood with a two Poisson distributions and a Gaussian systematic uncertainties~(\ref{poissnobsnbgeps})  is compared with a frequentist approach using a marginal likelihood with the background with a known uncertainty~(\ref{lihelihoodsysstd}) in fig.~\ref{fig:all_limits}. 
\section{Signal flux estimation using simulations}
In the previous sections the signal was estimated as a number of events. However sometimes the result should be presented as some source quantity (flux in astrophysics measurements, cross-section in acceleration experiments for example). Simulation of the source provides conversion of the latter quantity to the number of the signal events using a simple proportion:
\begin{equation}
\frac{s}{f}=\frac{s_\text{sim}}{f_\text{sim}}
\end{equation}
Simulation, however, is usually affected by a systematic uncertainty:
\begin{equation}
\mathcal{L}(s'_\text{sim})=\text{Gaussian}(s'_\text{sim}\mid s_\text{sim},\sigma_\text{sim})
\end{equation}
where the systematic uncertainty of the simulation is $\sigma_\text{sim}$. The uncertainty as a ratio $\sigma_\text{sim}/s_\text{sim}$ is equivalent to the detection efficiency uncertainty in~\citep{Conrad}.

The likelihood for $f$ can be written using the equation for signal $s$:
\begin{equation}
s=s_\text{sim}\frac{f}{f_\text{sim}}
\end{equation}
and~(\ref{lihelihoodsysstd}) in case of the classical measurements. The obtained likelihood is:
\begin{multline}
 \mathcal{L}(n_\text{obs}, b',s'_\text{sim} \mid f;b,f_\text{sim},\sigma_b,s_\text{sim},\sigma_\text{sim}) =\\
 = \text{Poisson}(n_{obs}\mid f\frac{s'_\text{sim}}{f_\text{sim}}+b')\text{Gaussian}(b'\mid b, \sigma_{b})\text{Gaussian}(s'_\text{sim}\mid s_\text{sim},\sigma_\text{sim})
\label{lihelihoodsysstdflux}
\end{multline}
 or for the case of the off-zones measurements using~(\ref{poissnobsnbgeps}):
 \begin{multline}
 \mathcal{L}(n_\text{obs},n_ \text{bg},\alpha_\text{on},\alpha_\text{off},s'_\text{sim} \mid f,b;f_\text{sim},\tau,\sigma,s_\text{sim},\sigma_\text{sim}) = \\
=\text{Poisson}(n_\text{obs} \mid \alpha_\text{on}b+f\frac{s'_\text{sim}}{f_\text{sim}})\text{Gaussian}(\alpha_\text{on}\mid1,\sigma)\times\\
\times \text{Poisson}(n_\text{bg} \mid \alpha_\text{off}\tau b)\text{Gaussian}(\alpha_\text{off}\mid 1, \sigma)\times\\
\times\text{Gaussian}(s'_\text{sim}\mid s_\text{sim},\sigma_\text{sim})
\label{poissnobsnbgepsflux}
 \end{multline}
 
\begin{figure}
\begin{center}
\includegraphics[width=0.8\textwidth]{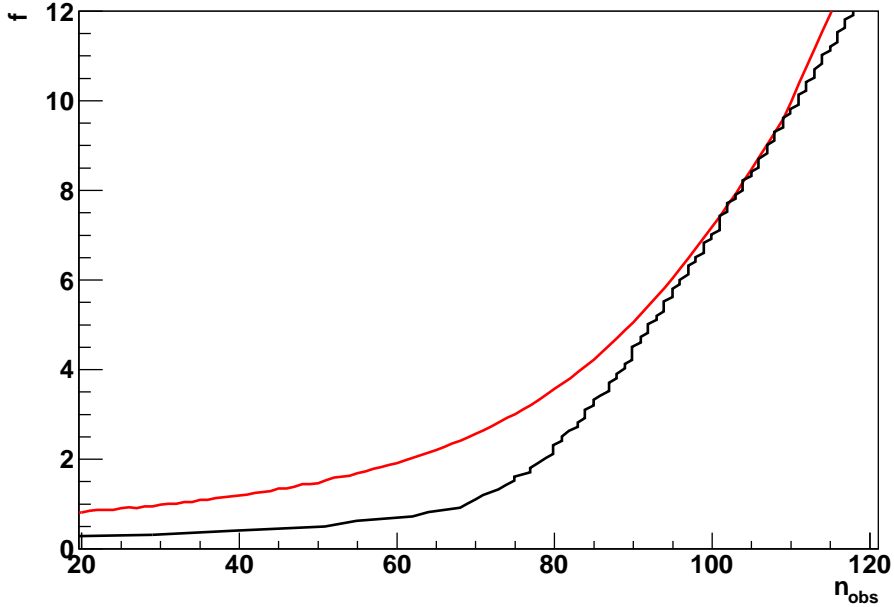}
\caption{Calculation of the upper limits with the likelihood~(\ref{lihelihoodsysstdflux}) for $b=90$, $\sigma_b=6$, $f_\text{sim}=1.2$, $s_\text{sim}=5.4$, $\sigma_\text{sim}=20\%$ with the Conrad's method (black) and with the likelihood~(\ref{poissnobsnbgepsflux}) for $n_\text{bg}=270$, $\tau=3$ and $\sigma=3\%$, $f_\text{sim}=1.2$, $s_\text{sim}=5.4$, $\sigma_\text{sim}=20\%$ with the Bayesian method using a profile likelihood (red).}
\label{fig:conrad_2gaus_flux}
\end{center}
\end{figure}

 The likelihood~(\ref{lihelihoodsysstdflux}) allows a marginalization and a frequentists limits construction which can be followed with minimal changes by~\citep{Conrad}. The Bayes approach using a profile likelihood is also applicable. The likelihood~(\ref{poissnobsnbgepsflux}) should be used with a Bayes approach as it depends from the two unknown parameters $s$ and $b$ (see discussions in section~\ref{sec:off}). The equations of the first derivatives for the profile likelihood parameters:
\begin{eqnarray}
\left(\frac{n_\text{obs}}{\alpha_\text{on}b+f\frac{s'_\text{sim}}{f_\text{sim}}}-1\right)\frac{f}{f_\text{sim}}-\frac{s'_\text{sim}-s_\text{sim}}{\sigma_\text{sim}^2}=0\\
\left(\frac{n_\text{obs}}{\alpha_\text{on}b+f\frac{s'_\text{sim}}{f_\text{sim}}}-1\right)\alpha_\text{on}+\left(\frac{n_\text{bg}}{\alpha_\text{off}\tau b}-1\right)\tau\alpha_\text{off}=0\\
\left(\frac{n_\text{obs}}{\alpha_\text{on}b+f\frac{s'_\text{sim}}{f_\text{sim}}}-1\right)b-\frac{\alpha_\text{on}-1}{\sigma^2}=0\label{form:s}\\
\left(\frac{n_\text{bg}}{\alpha_\text{off}\tau b}-1\right)\tau b-\frac{\alpha_\text{off}-1}{\sigma^2}=0\label{form:b}
\end{eqnarray}
where equation~(\ref{formula:alpha}) can be obtained again from the last three equations. The parameters $\hat\alpha_\text{on},\hat\alpha_\text{off},\hat s'_\text{sim},\hat b$ can be found by scanning $\alpha_\text{off}$ and obtaining $\hat b$ from equation~(\ref{form:b}), $\hat\alpha_{on}$ from equation~(\ref{formula:alpha}) and $\hat s'_\text{sim}$ from~(\ref{form:s}). Zero value should be used instead of the negative result for any of these parameters. The obtained limits calculated with both likelihoods are demonstrated in fig.~\ref{fig:conrad_2gaus_flux}.
\section{Significance estimation.}
To demonstrate the significance of the signal discovery the positive upper limit estimation may not be enough. Traditionally, an estimation how compatible is the measurement with the no signal hypothesis should be done also. If the process has a gaussian distribution with a mean $b$ and a sigma $\sigma$ and $n_\text{obs}$ was measured than the significance can be estimated simply as $S=(n_\text{obs}-b)/\sigma$ and one can say that "S~standard deviation was observed". For more complicated processes some test statistic which has a gaussian distribution for a background only hypothesis may be found.

This approach for the on/off-zone measurements is shown in~\citep{LiMa}. Basically it is done with a model hypothesis testing using a likelihood ratio test:
\begin{equation}
\lambda=\frac{\mathcal{L}(n_\text{obs},n_\text{bg} \mid s=0,\hat{b})}
{\mathcal{L}(n_\text{obs},n_\text{bg} \mid \hat{s},\hat{b})}
\end{equation}
where the likelihood maximum for the no signal hypothesis is found for the numerator and the best likelihood for all hypotheses is found for the denominator. The value of $-2log\lambda$ has approximately $\chi^2$ distribution for the no signal true model~\citep[chap. 10.3.1]{CasellaBerger}. Therefore, $\sqrt{-2log\lambda}$ has approximately a gaussian distribution and may be used as a significance estimation.

\begin{figure}
\begin{minipage}{0.5\textwidth}
\includegraphics[width=0.95\textwidth]{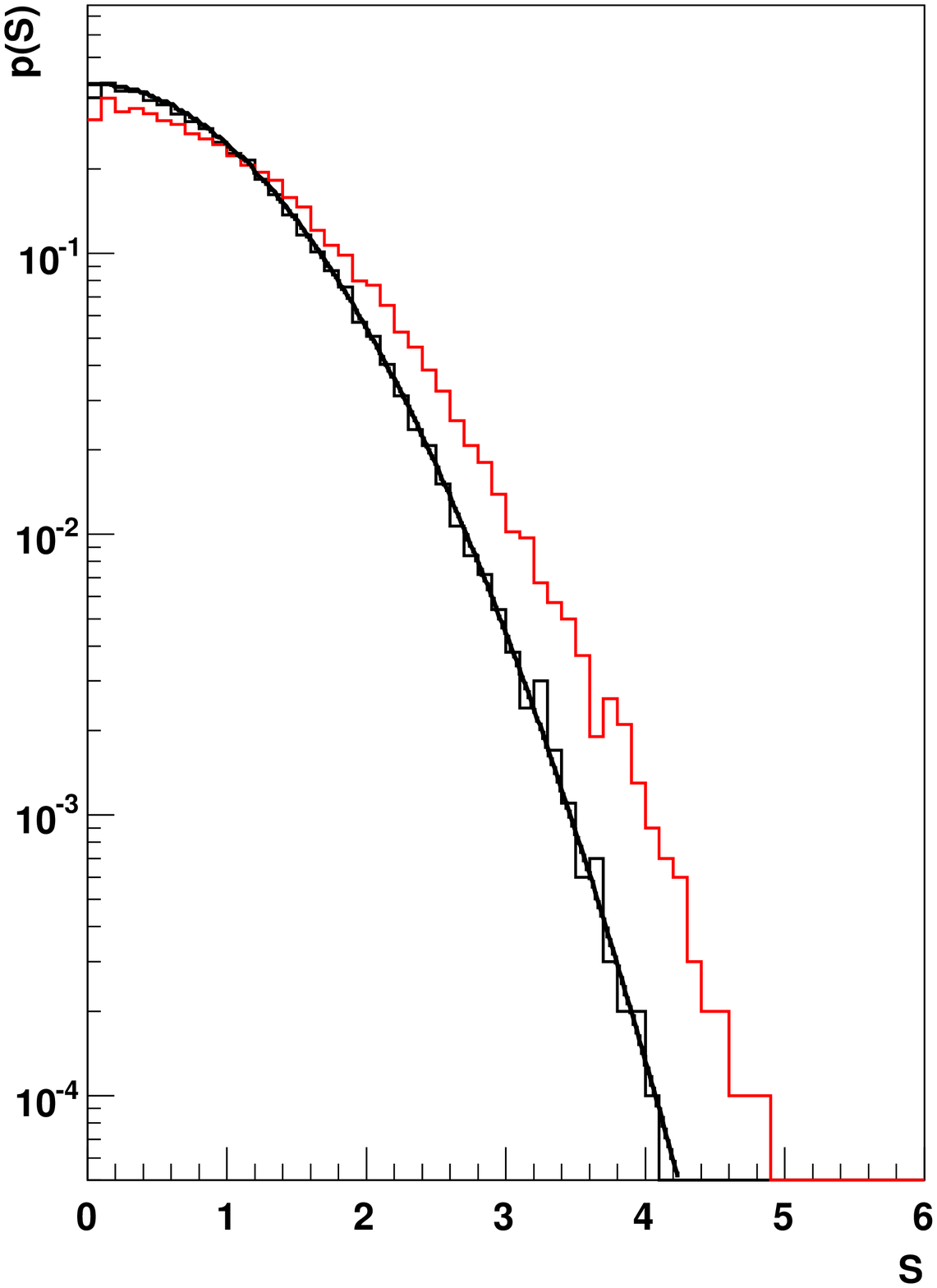}
\end{minipage}
\begin{minipage}{0.5\textwidth}
\includegraphics[width=0.95\textwidth]{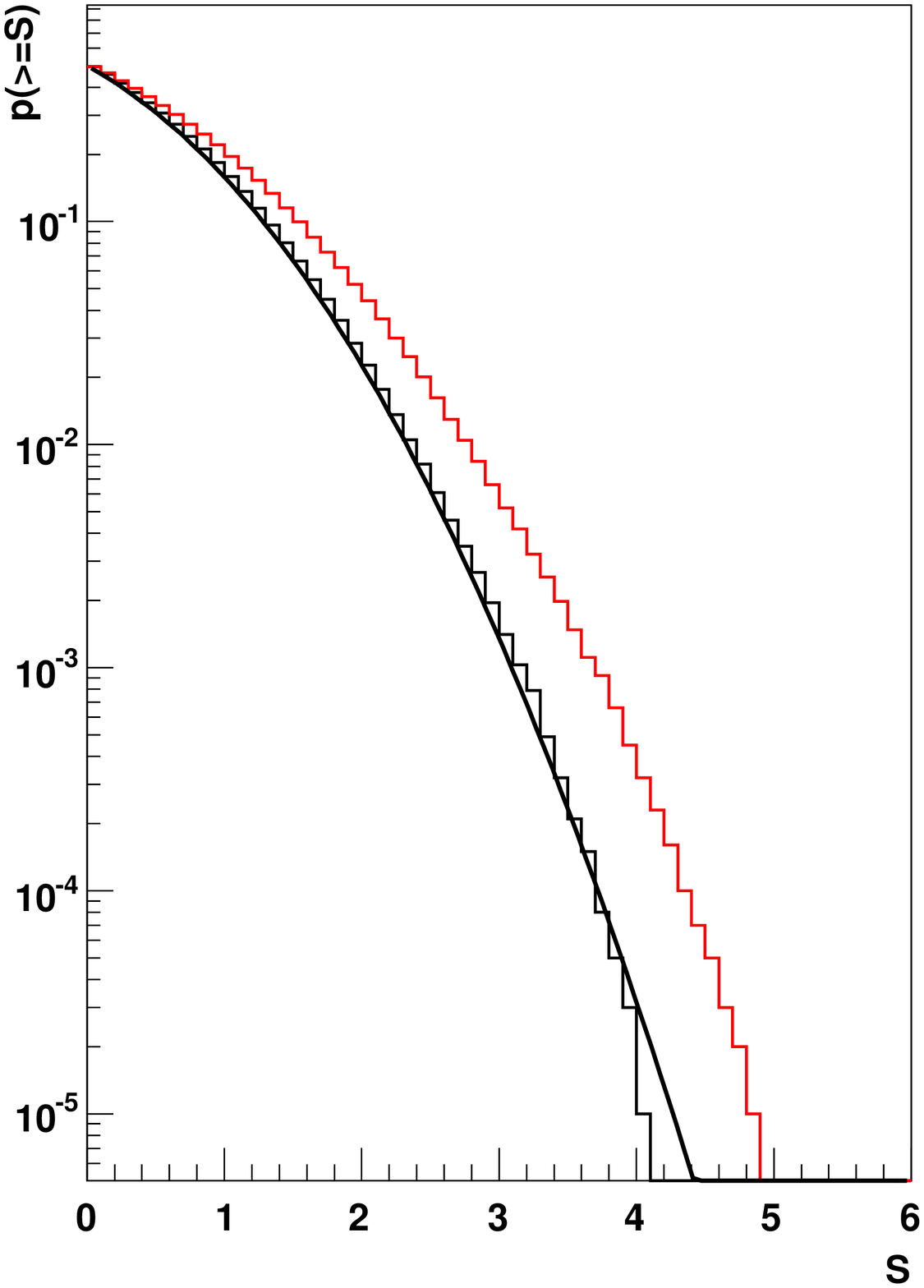}
\end{minipage}
\caption{Significance calculated for the Toy Monte Carlo simulations of the on/off-zone measurements with the mean background $b=90$, $\tau=3$ and the systematic uncertainty $\sigma=3\%$ for the model with an absence of the signal (black) compared to the calculations without a systematic uncertainty~\citep{LiMa} (red) and a gaussian distribution (smooth black line). }
\label{f:sign}
\end{figure}

For the off-zones method with systematic errors the  same procedure was applied. To find $\mathcal{L}(n_\text{obs},n_\text{bg} \mid s=0,\hat{b})$ or $\mathcal{L}(n_\text{obs},n_\text{bg} \mid \hat{s}\hat{b})$ a system of equations based on the fact that partial derivatives are zero at the local maximum was used. It was found for the denominator:
\begin{equation}
\hat b=\frac{n_\text{off}}{\alpha_\text{off}\tau};\,\,\,\,\,
\hat s=n_\text{on}-\hat \alpha_\text{on}\hat b;\,\,\,\,\,
\hat \alpha_\text{on}=1;\,\,\,\,\,
\hat \alpha_\text{off}=1;\,\,\,\,\,
\end{equation}

and for the numerator one can use equations~(\ref{formula:alpha}) and~(\ref{formula:b}) assuming s=0 and adopting the same scan procedure described in the previous section. The first equation remains the same and the last one can be solved as:
\begin{equation}
b_1=0,\,\,\,b_2=\frac{n_\text{on}+n_\text{off}}{\alpha_\text{on}+\alpha_\text{off}\tau}
\end{equation}
The obtained significance as a $\sqrt{-2log\lambda}$ was tested for a model with $n_{bg}=270$, $\tau=3$, $\sigma=3\%$ and compared with the significance from~\citep{LiMa} in fig.~\ref{f:sign} using a Toy Monte Carlo simulations. One can see that the calculated significance has an almost gaussian distribution.

\section{Conclusions}
In this note an overview of the existing methods for setting the upper limits with a presence of the unknown (nuisance) parameters  was done. The methods were used for the signal limits calculation for the case of the on/off-zones measurements. Also a new likelihood for this case of the measurements was constructed. It treats both observations in the on-zone and in the off-zone in the similar way and also includes the uncertainty of the expected background between the zones. 

It was shown on a particular example that all described methods with the profile and marginal likelihoods give the similar upper limits for the case when the number of the observed events is bigger than the expected background. If the number of the observed events is less than the background expectation then the frequentist methods tend to give smaller limits compared to the Bayesian methods.

The suggested method for the limits calculation in case of the on/off-zone measurements is the Bayesian approach together with the two Poisson distributions - one for the on-zone and another for the off-zone. It was shown also how to include the systematic uncertainties in this case. The significance of the signal presence using a background only hypothesis testing with a likelihood ratio test was presented for this likelihood as well. It was demonstrated that the significance distribution follows well the gaussian distribution with the mean 0 and the sigma 1.

\section{Acknowledgements}
I would like to thank Sotiris Loucatos, Gary C. Hill, Martin David Haigh, Juan Jos\'e Hern\'andez Rey, Domenico Costantini and Paschal Coyle for noticing the existing statistical methods and for the help in understanding of these methods as well.

\end{document}